\documentclass{article}

\usepackage{arxiv}

\usepackage[utf8]{inputenc} 
\usepackage[T1]{fontenc}    
\usepackage[unicode=true]{hyperref}
\usepackage{booktabs}
\usepackage{url}            
\usepackage{booktabs}       
\usepackage{amsfonts,amsmath,amssymb}       
\usepackage{nicefrac}       
\usepackage{microtype}      
\usepackage{cleveref}       
\usepackage{graphicx}
\usepackage[numbers]{natbib}
\usepackage{doi}

\title{Rethinking Thematic Evolution in Science Mapping:\\An Integrated Framework for Longitudinal Analysis\thanks{The algorithm underlying the proposed methodology has been implemented in the development version of the \texttt{R} package \texttt{bibliometrix}. The package is available through the project's GitHub repository: \href{https://github.com/massimoaria/bibliometrix}{https://github.com/massimoaria/bibliometrix}.}}

\usepackage{authblk}

\newbox{\orcid}\sbox{\orcid}{\includegraphics[scale=0.1]{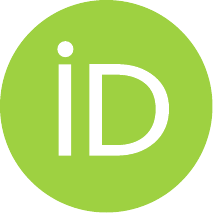}} 
\author[1]{Massimo Aria\hspace{1mm}%
	\href{https://orcid.org/0000-0002-8517-9411}{\usebox{\orcid}}\thanks{Corresponding author: \texttt{aria@unina.it}}%
}
\author[1]{Luca D'aniello\hspace{1mm}%
	\href{https://orcid.org/0000-0003-1019-9212}{\usebox{\orcid}}%
}
\author[2]{Michelangelo Misuraca\hspace{1mm}%
	\href{https://orcid.org/0000-0002-8794-966X}{\usebox{\orcid}}%
}
\author[1]{Maria Spano\hspace{1mm}%
	\href{https://orcid.org/0000-0002-3103-2342}{\usebox{\orcid}}%
}
\affil[1]{Department of Economics and Statistics, University of Naples Federico II, 81026 -- Naples, Italy}
\affil[2]{Department of Management \& Innovation Systems, University of Salerno, 84084 -- Fisciano, Italy}

\renewcommand{\shorttitle}{Rethinking Thematic Evolution in Science Mapping}

\hypersetup{
pdftitle={Rethinking Thematic Evolution in Science Mapping:\\An Integrated Framework for Longitudinal Analysis},
pdfsubject={cs.DL, cs.SI},
pdfauthor={Massimo Aria, Luca D'Aniello, MIchelangelo Misuraca, Maria Spano},
pdfkeywords={Co-word analysis, Fuzzy affiliation, Longitudinal network modelling},
}

\begin{document}

\maketitle

\begin{abstract}
Strategic diagrams and co-word analysis are widely employed to examine the conceptual structure of scientific domains and their development over time. Yet a structural inconsistency characterises dominant longitudinal implementations: themes are detected through relational clustering in weighted networks, whereas their inter-temporal connections are commonly inferred from set-theoretic overlap among keywords or core documents. This study introduces a structurally integrated framework in which lineage reconstruction is embedded within the same weighted relational architecture that underpins cross-sectional detection. The approach models thematic continuity through graded document affiliation and a lineage-strength measure that combines directional coverage with centrality-weighted structural relevance, thereby conceptualising evolution as the reconfiguration of relational structures rather than simple lexical persistence. By aligning thematic detection and temporal modelling within a unified relational paradigm, the framework enhances the methodological coherence and interpretive robustness of longitudinal science mapping.
\end{abstract}

\keywords{Co-word analysis\and Fuzzy affiliation\and Longitudinal network modelling}

\section{Introduction}\label{sec:1}
The quantitative study of scientific knowledge has increasingly relied on formal models that represent its structural organisation and temporal dynamics. Within this broader landscape, science mapping has emerged as a central methodological paradigm, complementing performance analysis by focusing on the relational architecture of research domains \citep{noyons1999integrating,borner2010atlas,vanraan2019measuring}. By leveraging bibliographic data, science mapping seeks to depict in a systematic and reproducible manner the intellectual and conceptual configuration of scientific fields. Among the available approaches, co-word analysis has played a pivotal role in operationalising conceptual structure. Since the seminal contributions of \citet{callon1983words} and the subsequent formalisation of the strategic diagram \citep{callon1991techno}, networks of term co-occurrence have been used to identify thematic areas and to characterise their structural role within a domain. The combination of community detection and the centrality-density framework has provided an interpretable representation in which themes are classified according to their strategic importance and internal development. Over time, this paradigm has been consolidated through methodological refinements and dedicated software tools \citep{cobo2012scimat,aria2017bibliometrix}, becoming a standard instrument in both cross-sectional and longitudinal bibliometric investigations.

The progressive growth of scientific production and the increasing interdisciplinarity of research have heightened interest in modelling the dynamic evolution of themes, moving beyond purely static configurations. Existing longitudinal extensions typically identify thematic clusters independently across successive time periods and subsequently establish intertemporal connections using measures of keyword overlap. This procedure has enabled the systematic detection of continuities, splits, and mergers, thereby offering valuable empirical insights into knowledge dynamics. In parallel, several alternative traditions have addressed longitudinal conceptual change through probabilistic topic models, embedding-based semantic drift measures, or dynamic community detection \citep{blei2006dynamic,hamilton2016diachronic,rossetti2018dynamic}. These approaches have expanded the methodological repertoire available for studying knowledge dynamics, yet they often involve trade-offs between interpretability, parameterisation, and the transparency of the linkage mechanism. In this setting, strategic mapping remains attractive because it provides an explicit structural representation of themes grounded in observable co-occurrence relations. The present work builds on this strength and targets a specific methodological discontinuity in dominant longitudinal implementations, namely the divergence between relational theme detection and predominantly lexical lineage construction. Yet the prevailing approach implicitly conceptualises thematic evolution as a problem of alignment between independently derived partitions. Continuity is generally inferred from lexical similarity, while the relational structure that defines themes cross-sectionally plays only a limited role in modelling their transformation. Moreover, document-level affiliation is not explicitly incorporated into lineage mechanisms, and thematic assignment remains predominantly crisp, notwithstanding the increasingly hybrid character of contemporary research. As a result, the representation of evolution may privilege the persistence of vocabulary over the preservation and reconfiguration of structural relations.

This paper proposes a re-specification of longitudinal thematic analysis grounded in a structurally integrated framework. Building upon the co-word tradition, the proposed approach reconceptualises thematic evolution as a network-based process in which cross-sectional configuration, graded document affiliation, and inter-temporal linkage are jointly modelled within a coherent analytical architecture. The objective is not to replace the strategic diagram paradigm, but to extend its explanatory scope by embedding thematic continuity in a formally defined relational representation that enhances methodological coherence and interpretability. 

\section{Conceptual and Methodological Background}\label{sec:2}
Co-word analysis is a foundational methodology for exploring the conceptual organisation of scientific domains. Introduced by \citet{callon1983words, callon1986mapping}, the approach rests on the premise that recurrent co-occurrence patterns among keywords reveal latent thematic structures. Terms are represented as nodes in a network, and their joint appearances define weighted edges; cohesive subgraphs extracted from this network are interpreted as themes. In this perspective, thematic areas emerge from relational configurations grounded in observed associations, without relying on predefined classifications. A decisive theoretical refinement was the evaluation of thematic clusters along two structural dimensions: centrality and density. Centrality reflects the extent to which a cluster's connections with the remainder of the network, and therefore its strategic positioning within the broader domain. Density measures internal cohesion and indicates the degree of conceptual development a theme has achieved. These dimensions were operationalised in the strategic diagram \citep{callon1991techno}, which maps themes onto a two-dimensional space and enables a functional interpretation of their roles within the field. The enduring relevance of this representation lies in its capacity to translate network-theoretic properties into an analytically meaningful typology \citep{law1992}.

Subsequent methodological developments have concentrated on improving the robustness of cross-sectional mapping. The adoption of association strength normalisation has been shown to attenuate frequency bias and preserve relational information in co-occurrence networks \citep{waltman2013systematic}. Other studies have demonstrated the sensitivity of thematic configurations to preprocessing choices, term selection, and thresholding decisions, highlighting the importance of methodological consistency. These refinements have strengthened the reliability of thematic detection within individual time slices. The extension of this paradigm to longitudinal analysis has generally proceeded by applying thematic detection independently to successive periods and then establishing connections among clusters across time. The framework proposed by \citet{cobo2011approach} systematised this procedure by introducing overlap-based similarity measures to identify continuities, splits, and mergers between themes. Inter-temporal relations are thus inferred from the degree of shared vocabulary between clusters identified in adjacent periods.

A structural asymmetry, however, exists in this dominant paradigm. Whereas cross-sectional detection is explicitly relational, lineage construction is typically reduced to set-theoretic comparisons among term lists, with the consequence that the weighted association structure defining a cluster’s semantic core does not enter the linkage mechanism. Continuity is thus inferred solely from lexical overlap, leaving aside the preservation or reconfiguration of underlying structural relations. Likewise, document-level dynamics are not incorporated into the definition of thematic evolution. Publications are indirectly associated with clusters via their constituent terms, yet their distribution across themes does not explicitly inform modelling of inter-temporal transitions. The reallocation of scientific production across evolving thematic configurations thus remains analytically external to the linkage procedure. The asymmetry is reinforced by the widespread reliance on crisp community detection, which executes mutually exclusive term assignment and only indirectly associates publications with themes through their vocabularies. Such partitions offer a clear decomposition of the term network, but they are not designed to represent graded thematic participation, which is increasingly salient in interdisciplinary environments in which both terms and publications span multiple domains.

These characteristics collectively indicate that existing longitudinal frameworks treat thematic evolution as a correspondence problem between independently derived partitions, generating a conceptual discontinuity between cross-sectional representation and temporal transformation. Addressing this discontinuity requires aligning thematic detection, document affiliation, and inter-temporal linkage within a unified relational architecture.

\section{An Integrated Framework for Longitudinal Thematic Analysis}\label{sec:3}
Let $\mathcal{D} = \{d_1, \ldots, d_n\}$ denote a set of publications and $T = \{t_1, \ldots, t_p\}$ the ordered set of time periods, with $t_i < t_{i+1}$. Each publication belongs to exactly one period on the basis of publication date, so that $\mathcal{D}^{(t)} \cap \mathcal{D}^{(t')} = \emptyset$ for $t \neq t'$ and $\bigcup_{t \in T} \mathcal{D}^{(t)} = \mathcal{D}$. For each period $t \in T$, let $\mathcal{K}^{(t)}$ denote the set of terms observed in $\mathcal{D}^{(t)}$. The analytical goal is twofold. First, we aim to identify for each period $t$ a set of thematic clusters representing coherent research topics:
\begin{equation}\label{eq:1}
    C^{(t)} = \{C_1^{(t)}, \ldots, C_h^{(t)}, \ldots, C_{n_t}^{(t)}\}
\end{equation}

Second, we aim to track the evolutionary relations between clusters in consecutive periods through a  lineage function:
\begin{equation}\label{eq:2}
    \mathcal{L}(C_h^{(t)}, C_{h'}^{(t+1)})    
\end{equation}

The resulting structure is a temporally ordered directed graph $\mathcal{G} = (V,E)$, where vertices are $V = \bigcup_{t \in T} C^{(t)}$ and edges represent statistically and substantively meaningful evolutionary connections. The strategy yields a formally coherent and interpretable account of thematic evolution based on weighted network structure and fuzzy publication affiliation, overcoming the limits of binary overlap heuristics.

\subsection{Cross-Sectional Thematic Representation}
The procedure begins by structuring and organising publications' content and performing thematic detection. For each period $t$, terms are extracted from $\mathcal{D}^{(t)}$ and a co-occurrence matrix $\mathbf{W}^{(t)} = [w_{ij}^{(t)}]$ is constructed. Terms can be listed in author keywords, index keywords, or $n$-grams extracted from titles or abstracts. The generic element of $\mathbf{W}^{(t)}$ is expressed as \textit{association} (or \textit{equivalence}) index \citep{peters1993,eck2009} between terms $k_i$ and $k_j$, defined as
\begin{equation}\label{eq:3}
    w_{ij}^{(t)} = \frac{c_{ij}^{(t)}}{\sqrt{c_{ii}^{(t)} c_{jj}^{(t)}}}    
\end{equation}
where $c_{ij}^{(t)}$ denotes the co-occurrence frequency of the terms, whereas $c_{ii}^{(t)}$ and $c_{jj}^{(t)}$ are their occurrences. The matrix $\mathbf{W}^{(t)}$ induces a weighted network on $\mathcal{K}^{(t)}$, capturing the semantic structure of the collection in period $t$. To obtain the cluster partition $C^{(t)}$ from the collection, a community detection algorithm is applied to $\mathbf{W}^{(t)}$ \citep{aria2024comparative}. For each cluster $C_h^{(t)}$, the set of related terms is denoted by $\mathcal{K}(C_h^{(t)}) \subseteq \mathcal{K}^{(t)}$. Structural properties are computed to depict and map clusters, such as centrality and density measures derived from the internal and external link configuration of $\mathbf{W}^{(t)}$ \citep{cobo2011approach}. Clusters with negligible cumulative term frequency are filtered out using a predefined threshold to ensure analytical stability.

\subsection{Fuzzy Publication-to-Cluster Assignment}
Hard partitioning at the publication level is replaced by a fuzzy membership scheme over the clusters $C^{(t)}$, considering that publications may exhibit varying degrees of affiliation with multiple themes. For each publication $d_i \in \mathcal{D}^{(t)}$ and cluster $C_h^{(t)}$, we compute a similarity score based on the overlap between the publication's terms and the cluster's characteristic terms, weighted by term centrality within the cluster, defined as
\begin{equation}\label{eq:4}
    s_{ih}^{(t)} = \sum_{k \in \mathcal{K}(d_i) \cap \mathcal{K}(C_h^{(t)})} 
\frac{\text{PR}_k^{(t)}(C_h)}{\text{freq}_k^{(t)}}    
\end{equation}
where $\text{PR}_k^{(t)}(C_h)$ is the \textit{PageRank centrality} \citep{brin1998,zhang2022} of term $k$ within cluster $C_h^{(t)}$ and $\text{freq}_k^{(t)}$ its period-specific frequency in $t$.

According to Eq.~\ref{eq:4}, publications' affiliation with a cluster is stronger when publications encompass more terms from the cluster itself, and these shared terms are central (high PageRank) within the cluster's semantic structure and distinctive (low frequency) rather than ubiquitous across the collection. The membership degree is then normalised to obtain a distribution over clusters
\begin{equation}\label{eq:5}
    u_{ih}^{(t)} = \frac{s_{ih}^{(t)}}{\sum_{j=1}^{n_t} s_{ij}^{(t)}},
\quad \text{with} \quad
\sum_{h=1}^{n_t} u_{ih}^{(t)} = 1
\end{equation}

The resulting fuzzy membership matrix $\mathbf{U}^{(t)} = [u_{ih}^{(t)}]$ encodes the multithematic nature of publications. In the rare case where a publication has zero similarity with all clusters (i.e., none of its terms appears in any cluster) due to a very restrictive term filtering, a uniform membership probability $u_{ih}^{(t)} = 1/n_t$ is assigned for all $h$ topics to ensure numerical stability. The effective size of each cluster, accounting for partial memberships, is computed as:
\begin{equation}\label{eq:6}
    |C_h^{(t)}| = \sum_{i=1}^{|\mathcal{D}^{(t)}|} u_{ih}^{(t)}    
\end{equation}
which captures cumulative and graded affiliations, avoiding the constraints of exclusive assignments.

\subsection{Inter-Temporal Assignment and Lineage Strength}
To formalise lineage, we distinguish between \textit{coverage} and \textit{structural relevance}. Coverage captures how much of a source theme's semantic mass is carried forward into a candidate successor, while structural relevance evaluates whether the shared content corresponds to conceptually central elements in both themes. These two dimensions are complementary: the former is directional and sensitive to what is retained from the source, whereas the latter reflects mutual thematic coherence. After performing period-specific analyses for all periods $t \in T$, we quantify the evolutionary connections between clusters in consecutive periods through a \textit{lineage strength} measure. This measure integrates two complementary dimensions: the weighted inclusion of terms and the importance of shared terms, measured by their centrality. Let $\mathcal{S}(C_h^{(t)}, C_j^{(t+1)}) = 
\mathcal{K}(C_h^{(t)}) \cap \mathcal{K}(C_j^{(t+1)})$ denote the shared terms between two periods. The \textit{weighted inclusion index} is defined as
\begin{equation}\label{eq:7}
    I_w(C_h^{(t)}, C_j^{(t+1)}) =
\frac{\sum_{k \in \mathcal{S}} \text{PR}_k^{(t)}(C_h)}
{\text{PR}_{tot}(C_h^{(t)})}  
\end{equation}
where $\text{PR}_{tot}(C_h^{(t)})$ is the total PageRank of cluster $C_h^{(t)}$. The index assumes values in $[0,1]$ and measures the proportion of cluster $C_h^{(t)}$'s total PageRank that is carried by terms also appearing in cluster $C_j^{(t+1)}$. It is important to highlight that the weighted inclusion index is asymmetric, so that $I_w(C_h^{(t)}, C_j^{(t+1)}) \neq I_w(C_j^{(t+1)}, C_h^{(t)})$. This asymmetry is meaningful because it reflects the extent to which the semantic content of the first cluster is retained in the second, while accounting for the relative importance of the shared terms.

The \textit{importance index} quantifies the aggregate relevance of shared terms by incorporating their centrality in both clusters through a normalised similarity measure, analogous to the association index in Eq.~\ref{eq:3}, but computed on PageRank values instead of co-occurrence frequencies. The index is defined as
\begin{equation}\label{eq:8}
    \Omega(C_h^{(t)}, C_j^{(t+1)}) = \sqrt{\frac{\sum_{k \in \mathcal{S}(C_h^{(t)}, C_j^{(t+1)})} \text{PR}_k^{(t)}(C_h) \cdot \text{PR}_k^{(t+1)}(C_j)}{\text{PR}_{tot}(C_h^{(t)}) \cdot \text{PR}_{tot}(C_j^{(t+1)})}}
\end{equation}

The index takes values in $[0,1]$ and captures the semantic coherence of the evolutionary connection by evaluating the centrality-weighted overlap between clusters. High values indicate that shared terms are not incidental lexical intersections but correspond to structurally central concepts within both thematic configurations. Conversely, low values suggest that the common vocabulary is peripheral to one or both clusters, signalling limited structural continuity. In contrast to the weighted inclusion index $I_w$, which is inherently asymmetric and measures the proportion of the source cluster’s semantic mass retained in the target cluster, the importance index is algebraically symmetric in its two arguments, being defined through a bilinear combination of PageRank centralities normalised by the product of their total masses. Nevertheless, because PageRank scores are computed within period-specific cluster subgraphs and may exhibit heterogeneous dispersion across periods, the empirical interpretation of this measure can display mild asymmetries. These do not stem from the index's functional form, but from differences in the underlying centrality distributions that condition the relative salience of shared terms in each temporal context. Joint consideration of $I_w$ and the importance index distinguishes alternative forms of intertemporal relatedness. When both directional coverage and mutual structural importance are high, the linkage reflects strong continuity: a substantial fraction of the source theme is transmitted, and the shared terms remain central in both configurations. High coverage combined with low importance indicates broad lexical retention without preservation of the semantic core. Conversely, low coverage paired with high importance captures selective transmission, whereby a limited set of shared terms -- yet highly central in both clusters -- anchors a focused but potentially substantive thematic connection. When both dimensions are low, the relation corresponds to weak continuity, characterised by marginal overlap and peripheral shared content. The integration of both measures in the lineage strength provides a comprehensive assessment of evolutionary connections that accounts for both the extent of overlap and the centrality of shared content
\begin{equation}\label{eq:9}
    \mathcal{L}(C_h^{(t)}, C_j^{(t+1)}) =
\alpha\, I_w(C_h^{(t)}, C_j^{(t+1)})
+ (1-\alpha)\, \Omega(C_h^{(t)}, C_j^{(t+1)})    
\end{equation}

The parameter $\alpha \in [0,1]$ regulates the emphasis placed on directional retention versus mutual structural relevance: values closer to one prioritise coverage of the source theme, values closer to zero prioritise the centrality-weighted coherence of the shared content in both themes, and $\alpha=0.5$ provides a balanced compromise. The parameter can be adjusted based on domain characteristics and analytical objectives. For each pair of consecutive periods $(t, t+1)$, a lineage strength matrix $\mathbf{L}^{(t,t+1)} = [\mathcal{L}_{hj}]$ is built, with a generic element equal to $\mathcal{L}(C_h^{(t)}, C_j^{(t+1)})$. The matrix provides a wide overview of all potential evolutionary connections among periods.

\subsection{Automatic Lineage Detection and Evolutionary Graph}
The final component of the approach automatically identifies significant evolutionary pathways and produces a thematic evolution graph for visualisation and interpretation. Since not all non-zero lineage strengths represent meaningful evolutionary connections, we apply a dual-thresholding approach that combines absolute and relative criteria to identify significant lineages. For each source cluster $C_h^{(t)}$, we identify target clusters in period $t+1$ that satisfy either of the following conditions
\begin{equation}\label{eq:10}
\mathcal{L}(C_h^{(t)}, C_j^{(t+1)}) \geq \theta_{\text{abs}}
\end{equation}
where $\theta_{\text{abs}}$ is a minimum lineage strength threshold, and
\begin{equation}\label{eq:11}
\text{rank}_h(C_j^{(t+1)}) \leq k
\end{equation}
where $\text{rank}_h(C_j^{(t+1)})$ denotes the rank of $C_j^{(t+1)}$ among all target clusters ordered by lineage strength from $C_h^{(t)}$. The dual criterion ensures that we capture both strongly connected components (exceeding the threshold) and relatively strong connections (top-k per source), respectively, preventing the loss of important evolutionary paths that may be weaker in absolute terms but represent the primary continuations of source clusters. The evolutionary graph $\mathcal{G} = (V, E, \mathbf{w})$ is directed and acyclic, where $V = \bigcup_{t \in T} C^{(t)}$ denotes the set of all thematic clusters identified across the entire temporal horizon, $E \subseteq V \times V$ is the set of directed edges $(C_h^{(t)}, C_j^{(t+1)})$ capturing significant evolutionary links between clusters in consecutive periods -- as determined by the adopted thresholding criteria -- and $\mathbf{w}: E \rightarrow [0,1]$ is a weight function assigning to each edge its corresponding lineage strength, such that $\mathbf{w}\big((C_h^{(t)}, C_j^{(t+1)})\big) = \mathcal{L}(C_h^{(t)}, C_j^{(t+1)})$.

The graph is temporally stratified: clusters from period $t$ form a layer $t$, and edges connect only consecutive layers. Evolutionary patterns are classified by the in-degree and out-degree of clusters in the resulting graph. A \textit{continuation} corresponds to a one-to-one link, with $|in(C^{(t+1)}_j)|=1$ and $|out(C^{(t)}_h)|=1$, indicating a stable trajectory across consecutive periods. A \textit{split} occurs when $|out(C^{(t)}_h)|>1$, suggesting differentiation or thematic specialisation, while a \textit{merge} occurs when $|in(C^{(t+1)}_j)|>1$, indicating thematic consolidation. A cluster is classified as \textit{emergent} when $|in(C^{(t+1)}_j)|=0$, and as \textit{disappearing} when $|out(C^{(t)}_h)|=0$, capturing, respectively, the appearance of a configuration without a significant predecessor and the absence of a significant successor. An evolutionary pathway is defined as a maximal directed sequence through the evolution graph using depth-first traversal
\begin{equation}\label{eq:12}
    \mathcal{P}_r = \langle C_{h_1}^{(t_1)}, \ldots, C_{h_q}^{(t_q)} \rangle    
\end{equation}
representing a coherent thematic trajectory spanning multiple periods. For each evolutionary pathway $\mathcal{P}_r = \langle C_{h_1}^{(t_1)}, \ldots, C_{h_q}^{(t_q)} \rangle$, we compute a set of aggregate indicators summarising its structural and temporal properties. Pathway strength is defined as the product of lineage strengths along consecutive transitions
\begin{equation}\label{eq:13}
    \text{Strength}(\mathcal{P}_r) = \prod_{i=1}^{q-1} \mathcal{L}\big(C_{h_i}^{(t_i)}, C_{h_{i+1}}^{(t_{i+1})}\big)    
\end{equation}
capturing the overall continuity of the trajectory. Pathway length corresponds to the number of periods spanned by the sequence, providing a measure of temporal persistence. Finally, cumulative size is defined as 
\begin{equation}\label{eq:14}
    \sum_{i=1}^{q} |C_{h_i}^{(t_i)}|    
\end{equation}
which aggregates the fuzzy cardinalities of the clusters that compose the pathway and reflects its overall substantive weight over time.

The evolutionary graph can be visualised through a thematic evolution plot structured along the temporal dimension. Time periods are arranged on the horizontal axis, while nodes represent thematic clusters identified in each period. The size of each node is proportional to the fuzzy cardinality $|C_h^{(t)}|$, thereby reflecting the substantive weight of the cluster. Directed edges connect clusters across consecutive periods, with edge thickness scaled according to the corresponding lineage strength $\mathcal{L}$, providing an immediate visual indication of the intensity of evolutionary connections. Distinct colours are used to differentiate evolutionary pathways, facilitating the identification of coherent thematic trajectories. The vertical positioning of nodes may encode additional structural information, such as centrality, density, or quadrant location in the strategic diagram, thereby integrating longitudinal and cross-sectional perspectives within a unified visual representation.

The framework is modular, allowing sensitivity analyses with respect to $\alpha$, threshold parameters, and clustering resolution. From a computational perspective, the dominant costs arise from community detection within each period-specific co-occurrence network and from the construction of inter-temporal lineage matrices. Let $p_t = |K^{(t)}|$ denote the number of retained terms and $m_t$ the number of non-zero edges in the weighted co-occurrence network $W^{(t)}$ for period $t$. Under standard sparsity conditions ($m_t \ll p_t^2$), modularity-based community detection via the Louvain algorithm operates in approximately $O(m_t)$ time per iteration, yielding near-linear empirical scaling in the number of edges. PageRank centrality, computed within cluster-induced subgraphs, has complexity $O(m_{th} I)$ for cluster $h$, where $m_{th}$ denotes its internal edges and $I$ the number of power iterations required for convergence, typically small in practice. Inter-temporal lineage construction requires pairwise comparison of clusters across adjacent periods. If $c_t = |C^{(t)}|$ denotes the number of clusters in period $t$, the lineage matrix between $t$ and $t+1$ involves at most $O(c_t c_{t+1} \bar{s})$ operations, where $\bar{s}$ represents the average number of shared terms per cluster pair. Since $c_t \ll p_t$ in typical applications, this component remains modest relative to network construction. Overall complexity can therefore be expressed as 
\begin{equation}\label{eq:15}
O\!\left(\sum_t m_t + \sum_t c_t c_{t+1} \bar{s} \right)    
\end{equation}
with the edge-based term $\sum_t m_t$ dominating under realistic sparsity. Importantly, scaling is governed by the number of retained terms and their relational density rather than by the total number of publications, ensuring feasibility for medium- and large-scale bibliometric collections.

\section{Empirical Application}\label{sec:4}
To illustrate the analytical potential of the proposed framework, the empirical application considers the complete publication record of the \textit{Journal of Informetrics} (JOI), one of the principal international outlets in scientometrics and bibliometrics. The journal constitutes an appropriate testbed given its thematic centrality and sustained development over nearly two decades. The dataset was retrieved from the Web of Science Core Collection and covers 2007--2025. After filtering for citable items, the final \textit{corpus} comprises 1{,}400 articles authored by 2{,}151 distinct scholars, with a mean of 2.87 co-authors per document and an international co-authorship rate of 30.07\%. Average citations per document equal 40.72 over the full window, and reference lists contain 32{,}929 cited items, indicating a dense intertextual structure. Author-assigned keywords number 3{,}887 unique terms (1{,}594 Keywords Plus), reflecting substantial semantic heterogeneity and conceptual evolution.

\begin{figure}[!ht]
\centering
\includegraphics[width=0.8\textwidth]{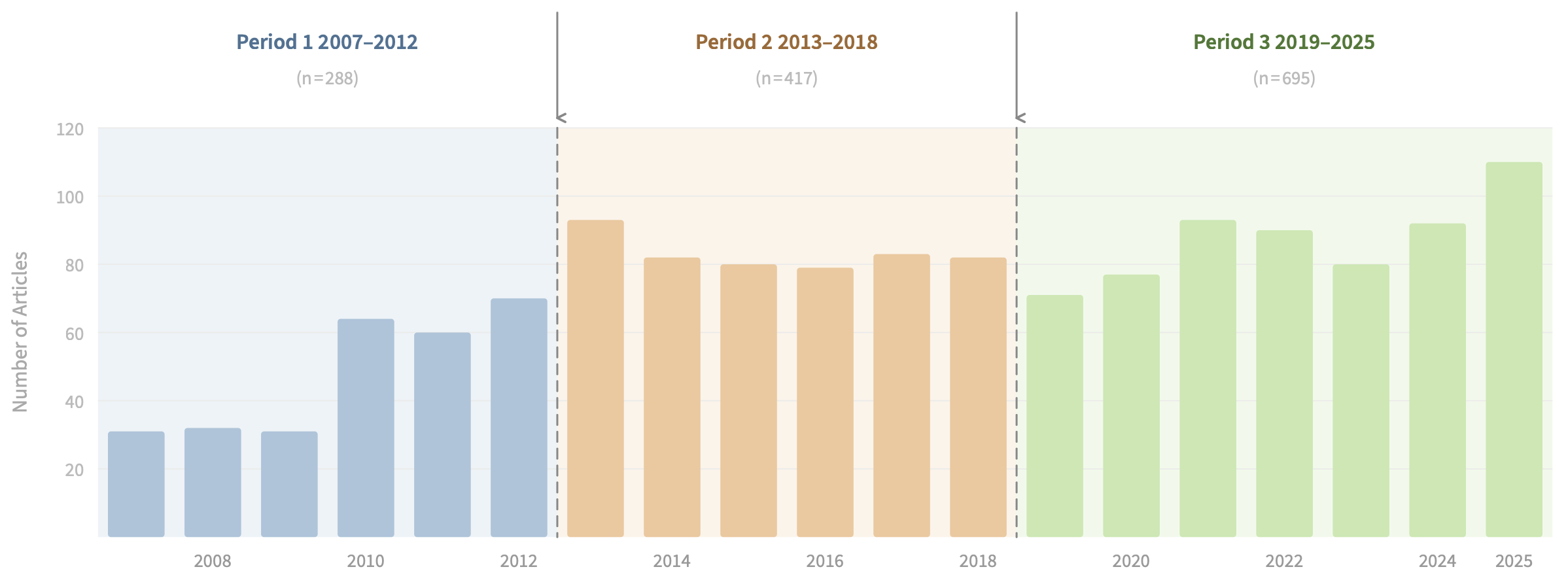}
\caption{Annual scientific production of the \textit{Journal of Informetrics} (2007--2025). Bars are coloured by analytical period; vertical grey lines indicate the two cutting points (end of 2012 and 2018), with period-level document counts shown within each shaded segment.}
\label{fig:asp}
\end{figure}

Figure~\ref{fig:asp} reports annual production into three sub-periods -- 2007--2012, 2013--2018, and 2019--2025 -- comprising 288, 417, and 695 articles, respectively.

Output increases from 31 articles in 2007 to 110 in 2025, corresponding to an average annual growth rate of 7.29\%. Period~2 exceeds Period~1 by 44.8\%, and Period~3 is 66.7\% larger than Period~2, indicating sustained acceleration. The first cutting point (end of 2012) is substantively motivated by the publication of the \textit{San Francisco Declaration on Research Assessment}, which intensified debate on responsible evaluation and coincides with a visible rise in publication volume (70 articles in 2012; 93 in 2013). The second cutting point (end of 2018) reflects both the consolidation of altmetrics as a distinct research stream and the diffusion of the \textit{Leiden Manifesto} principles, whose influence becomes structurally evident by the close of the decade. The 2019--2025 period thus captures a phase of thematic consolidation and diversification.

Co-word analysis was conducted on author-assigned keywords as proxies for conceptual structure. For each period, a co-occurrence matrix was constructed and normalised using the association strength measure defined in Eq.~(\ref{eq:3}), thereby attenuating frequency effects and enabling cross-period comparisons. A minimum threshold of five occurrences was imposed, retaining up to 250 terms per period. Prior to network construction, 280 pairs of semantically equivalent terms were harmonised (e.g., \textit{citation}/\textit{citations}, \textit{bibliometrics}/\textit{bibliometric}) to reduce artificial fragmentation. Community detection was performed using the \textit{Louvain} algorithm \citep{blondel2008}, which was selected for its efficiency and performance in weighted networks \citep{yang2016}. The resulting partitions provide the basis for fuzzy membership estimation and lineage reconstruction. Inter-period lineage strengths were computed as specified above, with $\alpha = 0.5$ to balance directional coverage and reciprocal structural relevance.

\subsection{Cross-Sectional Thematic Structure}
\begin{figure}[!ht]
\centering
\includegraphics[width=0.75\textwidth]{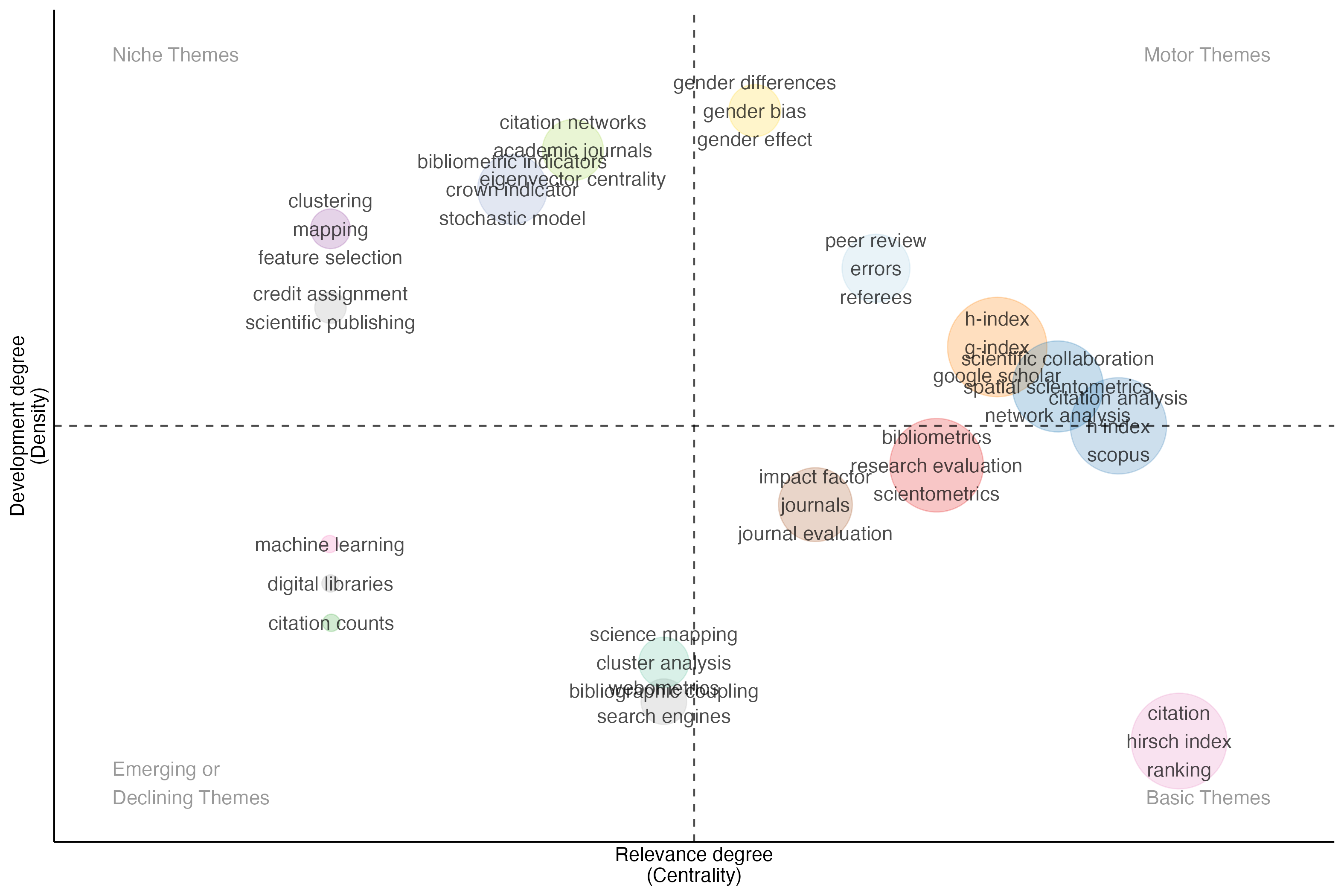}
\caption{Strategic Diagram: Period 1 (2007--2012).}
\label{fig:map1}
\end{figure}

\begin{figure}[!ht]
\centering
\includegraphics[width=0.75\textwidth]{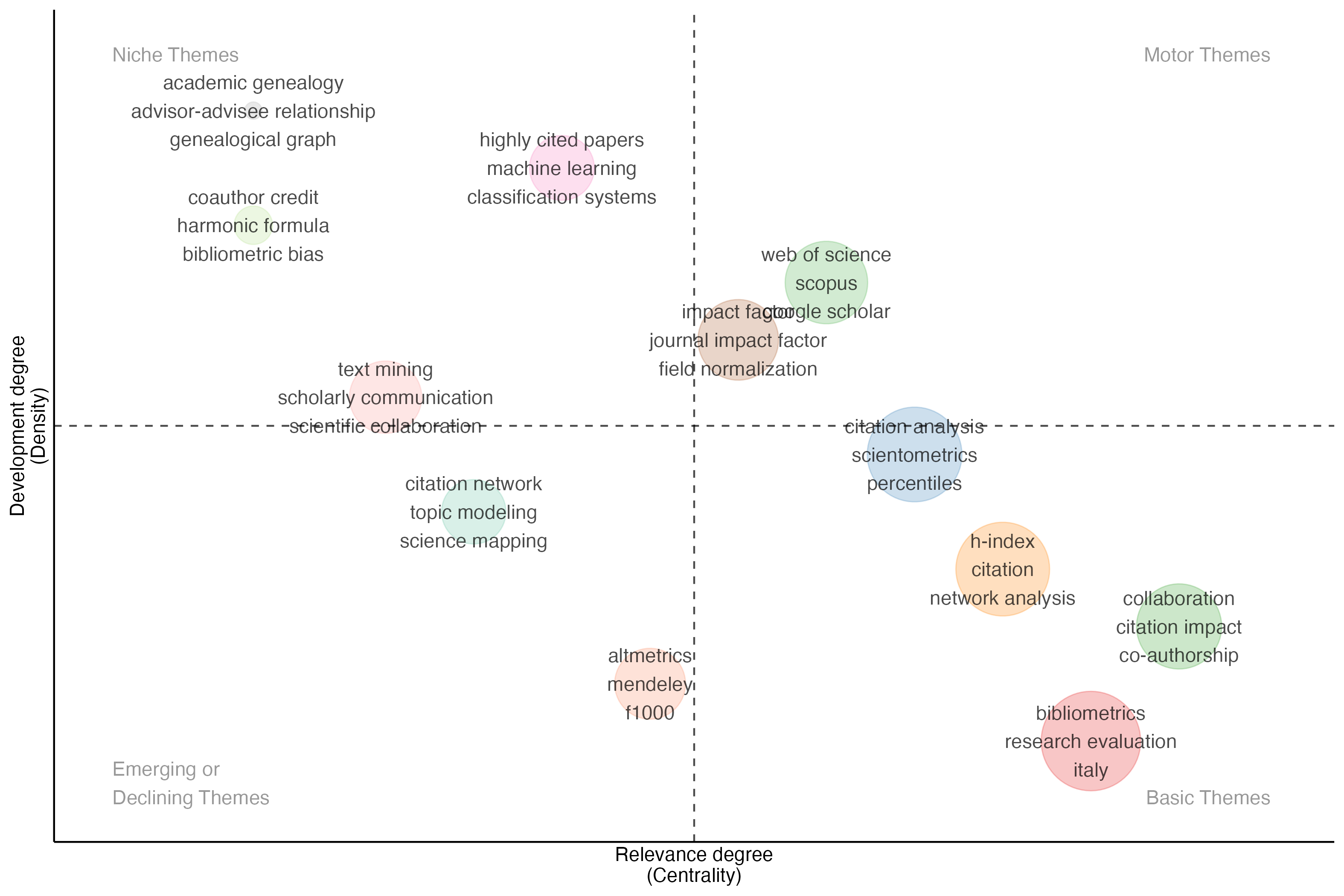}
\caption{Strategic Diagram: Period 2 (2013--2018).}
\label{fig:map2}
\end{figure}

\begin{figure}[!ht]
\centering
\includegraphics[width=0.75\textwidth]{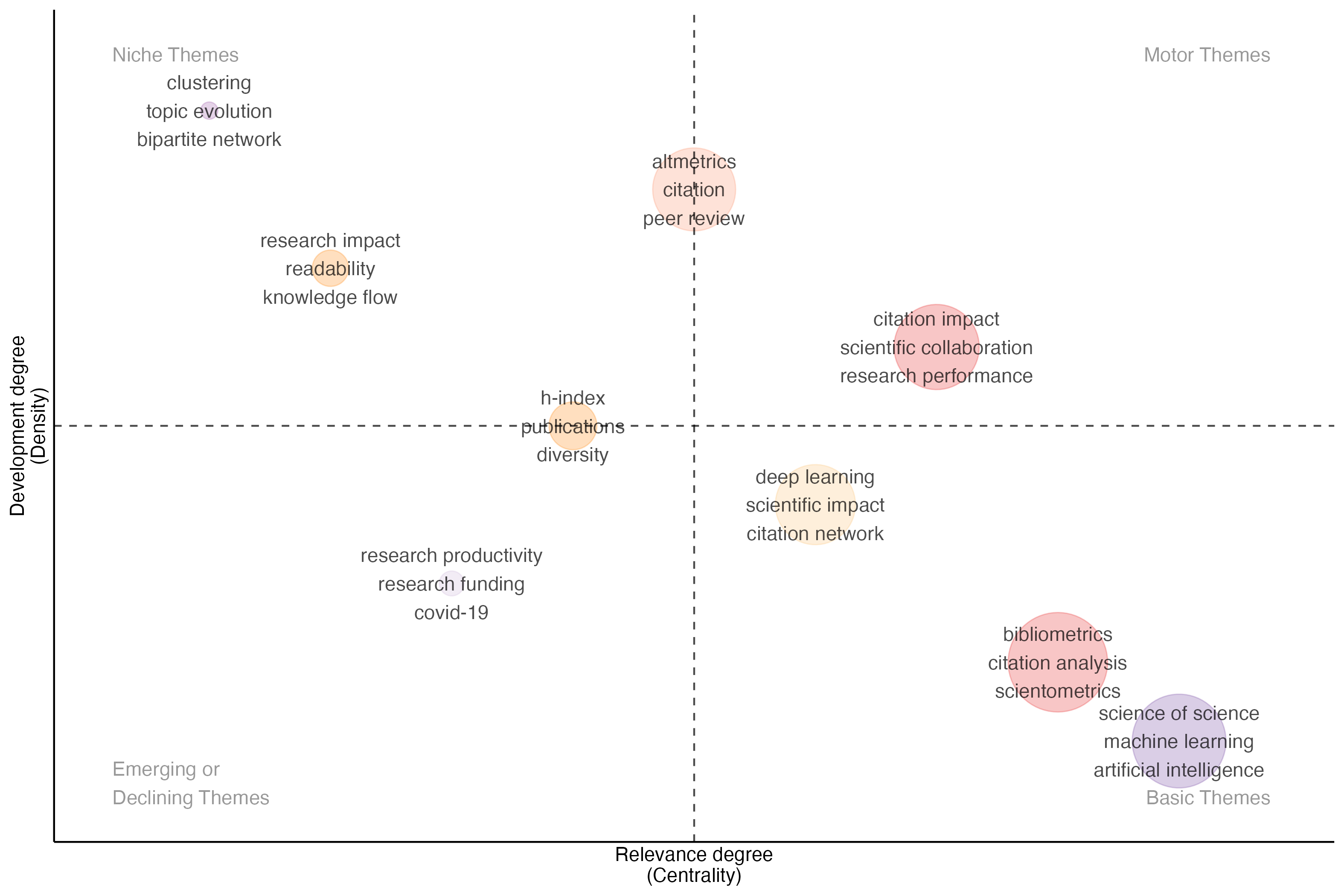}
\caption{Strategic Diagram: Period 3 (2019--2025).}
\label{fig:map3}
\end{figure}

The cross-sectional analysis identifies 18 thematic clusters in Period~1, 12 in Period~2, and 9 in Period~3. This progressive reduction in the number of clusters is not indicative of thematic contraction; rather, it reflects increasing structural consolidation and higher internal density within the co-word networks over time. As the field matures, previously fragmented or weakly connected thematic areas tend to coalesce into more cohesive and internally articulated research streams, leading to fewer but structurally stronger communities. The strategic diagrams corresponding to each period are presented in Figures~\ref{fig:map1}--\ref{fig:map3}. These visualisations provide a comparative representation of cluster centrality and density, enabling the identification of motor themes, basic and transversal themes, emerging or declining areas, and highly developed but isolated domains within each temporal configuration.

\paragraph{Period 1 (2007--2012)} The thematic configuration of JOI in its founding phase is structured around a diversified yet methodologically cohesive core centred on bibliometric indicators. The most frequent themes include \textit{h-index} (112 occurrences), \textit{citation analysis} (96), \textit{citation} (93), and \textit{bibliometrics} (84). In the strategic diagram, \textit{h-index} is located in the upper-right quadrant (motor themes), combining high centrality and high density, indicating a well-developed, structurally integrated research stream. \textit{Citation} is located in the lower-right quadrant, where its very high centrality, combined with more moderate density, characterises it as a basic and transversal theme that integrates diverse areas of the domain, not as a self-contained subfield. Related constructs such as \textit{scientific collaboration} and \textit{peer review} also appear among the motor themes, reflecting the consolidation of evaluation practices and collaborative dynamics as core components of the journal's early intellectual profile. Themes including \textit{impact factor}, \textit{journals}, and \textit{journal evaluation} cluster within the right-hand side of the diagram, reinforcing the centrality of research assessment debates in this period. In contrast, \textit{clustering} and \textit{mapping} are located in the upper-left quadrant (niche themes), characterised by relatively high internal density but limited centrality, suggesting specialised methodological developments with restricted integrative influence. Finally, \textit{machine learning}, \textit{digital libraries}, and \textit{citation counts} appear in the lower-left quadrant, combining low density and low centrality, consistent with emerging or weakly institutionalised trajectories that had not yet attained structural prominence within the field.

\paragraph{Period 2 (2013--2018)} The intermediate period is characterised by a marked expansion in publication volume (+44.8\% relative to Period~1) and a visible reorganisation of the thematic structure. The clusters decrease to 12, while the dominant themes increase substantially in size: \textit{bibliometrics} (210 occurrences), \textit{citation analysis} (171), \textit{h-index} (164), \textit{collaboration} (113), \textit{web of science} (101), and \textit{impact factor} (93). In the strategic diagram, \textit{web of science}, \textit{impact factor}, and closely related evaluation constructs are positioned in the upper-right quadrant, indicating well-developed motor themes. By contrast, \textit{collaboration}, together with \textit{bibliometrics} and \textit{research evaluation}, occupies the lower-right quadrant, functioning as highly central but comparatively less dense basic themes that structure the intellectual core of the period. The \textit{h-index} and \textit{citation} cluster remains strongly central, consolidating its role as a transversal evaluative framework. A salient development is the emergence of \textit{altmetrics} (61 occurrences) as a distinct cluster located in the lower-left quadrant, signalling a trajectory that, while thematically innovative and increasingly visible in the literature, had not yet attained structural maturity. The cluster \textit{highly cited papers} (47) appears in the upper-left quadrant, suggesting a specialised and internally cohesive niche. At the same time, \textit{text mining} (65) and \textit{citation network} (46), together with \textit{topic modelling} and \textit{science mapping}, indicate a growing methodological diversification, though these remain positioned outside the motor core. Finally, \textit{academic genealogy} and related descriptors form a niche theme characterised by high density but limited centrality, reflecting a specialised line of inquiry with restricted integrative influence.

\paragraph{Period 3 (2019--2025)} 
The most recent period is characterised by further output expansion (+66.7\% relative to Period~2) and a pronounced structural consolidation, with nine clusters that are substantially larger and more internally articulated than in earlier phases. The thematic configuration reflects the combined influence of computational advances and a growing emphasis on systemic and societal dimensions of science. In the strategic diagram, \textit{bibliometrics} (222 occurrences) and \textit{science of science} (188) occupy the lower-right quadrant, functioning as highly central but comparatively less dense basic themes. Their position indicates a broad reorientation of the journal's intellectual core towards macro-level analyses of scientific systems, while maintaining strong integrative capacity across subdomains. A second major cluster in the basic quadrant is represented by \textit{machine learning} and \textit{artificial intelligence}, signalling the normalisation of computational approaches within the field. The upper-right quadrant (motor themes) is dominated by \textit{citation impact}, \textit{scientific collaboration}, and \textit{research performance}, which exhibit high centrality and density, indicating mature, structurally cohesive research streams. By contrast, \textit{altmetrics}, \textit{citation}, and \textit{peer review} are located near the vertical axis with relatively high density but more moderate centrality, suggesting well-developed yet not fully transversal clusters. Methodological specialisation persists in the upper-left quadrant, where \textit{clustering}, \textit{topic evolution}, and \textit{bipartite network} form niche themes characterised by strong internal cohesion but limited integrative reach. Emerging or context-specific topics such as \textit{research productivity}, \textit{research funding}, and \textit{covid-19} appear in the lower-left quadrant, reflecting trajectories that are either newly consolidated or episodically driven. Finally, the \textit{h-index} cluster, though still present (49 occurrences), is positioned closer to the central boundary and shows reduced prominence compared with earlier periods, indicating a gradual rebalancing of attention away from classical indicator-centred research.

\subsection{Thematic Evolution}
\begin{figure}[!ht]
\centering
\includegraphics[width=0.85\textwidth]{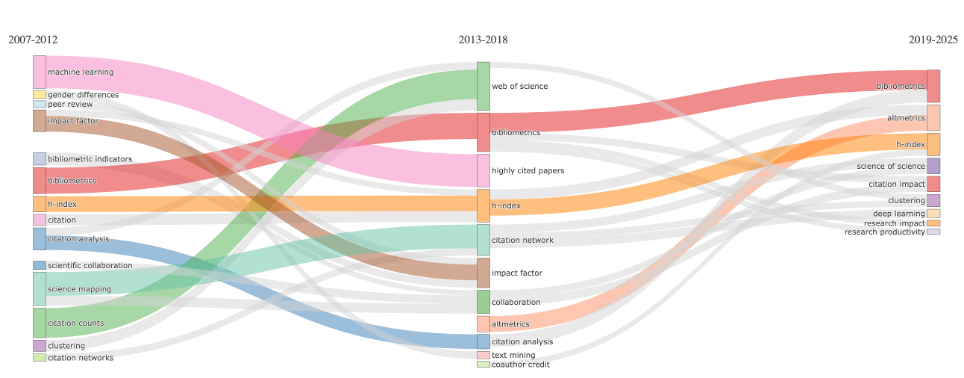}
\caption{Evolutionary graph.}
\label{fig:sankey}
\end{figure}

The evolutionary graph produced by the proposed framework, shown in Figure~\ref{fig:sankey}, maps the lineage relations among clusters across the three periods. Lineage strengths are computed using the integrated measure defined in Eq.~(\ref{eq:9}) with $\alpha = 0.5$, thereby balancing directional coverage and reciprocal structural relevance.

The most robust longitudinal trajectory concerns the \textit{bibliometrics} cluster, which exhibits the strongest continuation across both transitions. The flow from Period~1 to Period~2 is substantial, and the subsequent transition to Period~3 remains one of the dominant streams in the diagram. This continuity reflects the persistent integrative role of bibliometric methodology within the journal's intellectual structure. A similarly stable pathway characterises \textit{citation analysis}, which maintains visible connections across periods, preserving its semantic focus on indicator construction and scientometric modelling. The \textit{h-index} cluster also follows a continuous trajectory, although with progressively reduced magnitude in the final period, indicating relative contraction without complete disappearance. Between Period~1 and Period~2, the Sankey structure reveals a marked diversification of the highly central \textit{citation} theme. This latter evolves from a single dominant cluster into multiple successor themes, distributing its weight across \textit{h-index}, \textit{citation analysis}, and, to a lesser extent, \textit{altmetrics}, thereby reflecting the specialisation of citation-based inquiry into distinct evaluative and metric-oriented strands. Concurrently, \textit{science mapping} and \textit{scientific collaboration} feed into \textit{citation network} and \textit{collaboration}, illustrating a reorganisation of network-analytic approaches. The transition from Period~2 to Period~3 is characterised more by convergence than fragmentation. Notably, \textit{collaboration}, \textit{citation network}, and parts of \textit{h-index} and \textit{citation analysis} contribute to the emergence of the \textit{science of science} cluster, which appears in the final period as a major basic theme. This pattern indicates an integrative shift towards macro-level analyses of knowledge systems. In parallel, computational strands consolidate: early \textit{machine learning} and \textit{text mining} components feed into \textit{deep learning}, which emerges in Period~3 as a distinct cluster linked to citation-network methodologies. The trajectory of \textit{altmetrics} is particularly illustrative. Absent as a named cluster in Period~1, it appears in Period~2 and then splits in the subsequent transition, contributing both to a continued \textit{altmetrics} cluster and to \textit{citation impact} in Period~3. This dual continuation reflects partial institutionalisation alongside integration into broader impact-oriented frameworks. 


To assess the robustness of lineage reconstruction to the weighting parameter, the analysis was repeated with moderate perturbations of $\alpha$ (0.3 and 0.7). Observed differences are confined to weaker and peripheral connections. Lower values of $\alpha$ slightly emphasise structurally concentrated transmissions based on highly central shared terms, whereas higher values marginally strengthen broader directional carryover from larger source clusters. These adjustments affect the relative intensity of secondary links but do not modify the principal continuations, splits, or mergers, nor the overall macro-structure of the evolutionary graph. Although only the $\alpha = 0.5$ configuration is reported for parsimony, the alternative specifications yield an evolutionary topology that is substantively unchanged. In particular, the dominant longitudinal backbone is preserved across settings: \textit{bibliometrics} remains the strongest and most stable trajectory over both transitions; \textit{h-index} exhibits continuity with gradual attenuation; and the convergence of \textit{collaboration}, \textit{citation network}, and adjacent evaluative strands into \textit{science of science} in the final period is consistently reconstructed. The consolidation of \textit{altmetrics} between Period 2 and Period 3 likewise persists under both alternative values. The results, therefore, indicate that the main evolutionary patterns are not artefacts of a specific parameter choice, while confirming that $\alpha$ meaningfully regulates the balance between directional coverage and structural relevance in marginal transmissions.

\subsection{Comparative Assessment with Classical Thematic Evolution}
To contextualise the analytical contribution of the proposed framework, its outputs are compared with those generated by SciMAT \citep{cobo2012scimat}, the most widely adopted implementation of the classical longitudinal approach to thematic evolution. As before, the analysis was performed on the complete publication record of the \textit{Journal of Informetrics} (2007--2025), partitioned into three identical sub-periods: 2007--2012, 2013--2018, and 2019--2025. SciMAT was configured using \textit{words} (author keywords only) as the unit of analysis. Co-occurrence networks were normalised by association strength; clusters were extracted using single-pass centroid clustering (maximum cluster size set to 100); and the inclusion index was used as both an evolution and an overlap measure. This configuration mirrors the standard operationalisation of the Cobo framework and ensures procedural comparability. The comparison is structured along three dimensions: thematic granularity, cross-sectional cluster composition, and structure of inter-temporal lineage.

\begin{table}[ht]
\centering
\caption{Number of thematic clusters by period and approach.}
\label{tab:cluster_comparison}
\begin{tabular}{lccc}
\toprule
\textbf{Period} & \textbf{N. of doc.} & \textbf{Proposed framework} & \textbf{SciMAT} \\
\midrule
2007--2012 & 288 & 18 & ~7 \\
2013--2018 & 417 & 12 & 14 \\
2019--2025 & 695 & ~9 & 14 \\
\bottomrule
\end{tabular}
\end{table}

\begin{figure}[ht]
  \centering
  \includegraphics[width=0.5\textwidth]{%
    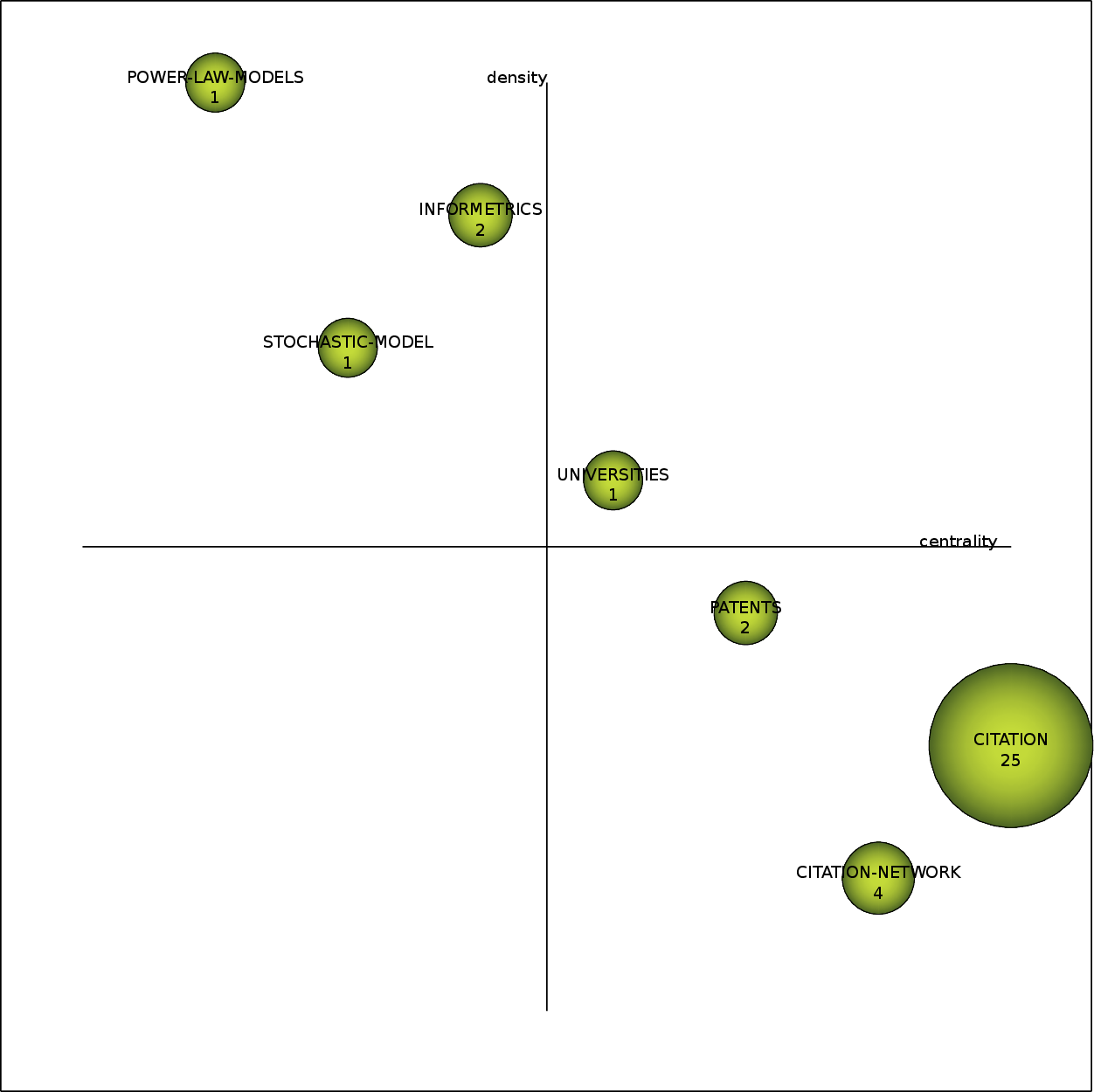}
  \caption{SciMAT strategic diagram:: Period~1 (2007--2012).}
  \label{fig:scimat_sd_p0}
\end{figure}

\begin{figure}[ht]
  \centering
  \includegraphics[width=0.5\textwidth]{%
    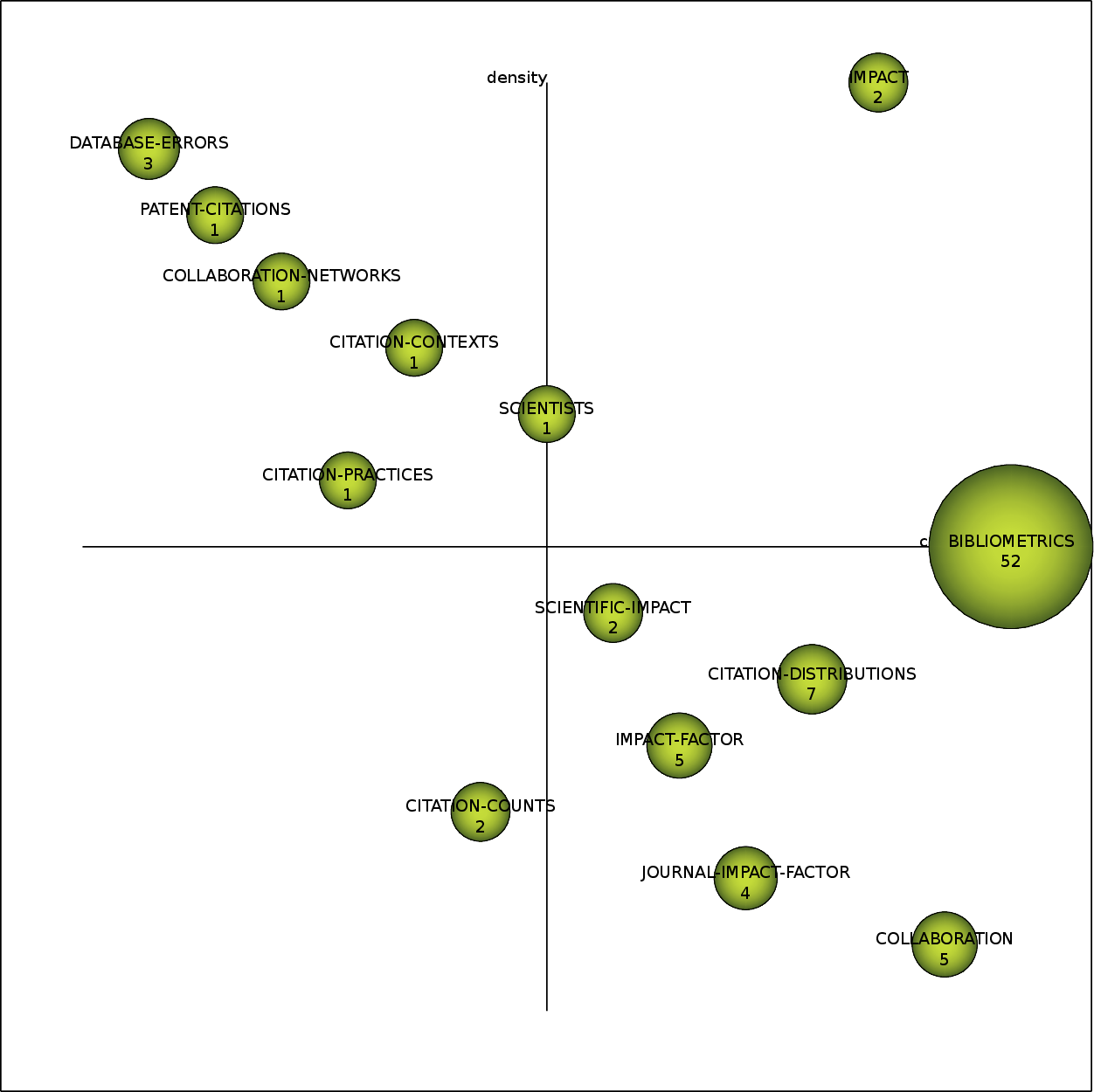}
  \caption{SciMAT strategic diagram: Period~2 (2013--2018).}
  \label{fig:scimat_sd_p1}
\end{figure}

\begin{figure}[ht]
  \centering
  \includegraphics[width=0.5\textwidth]{%
    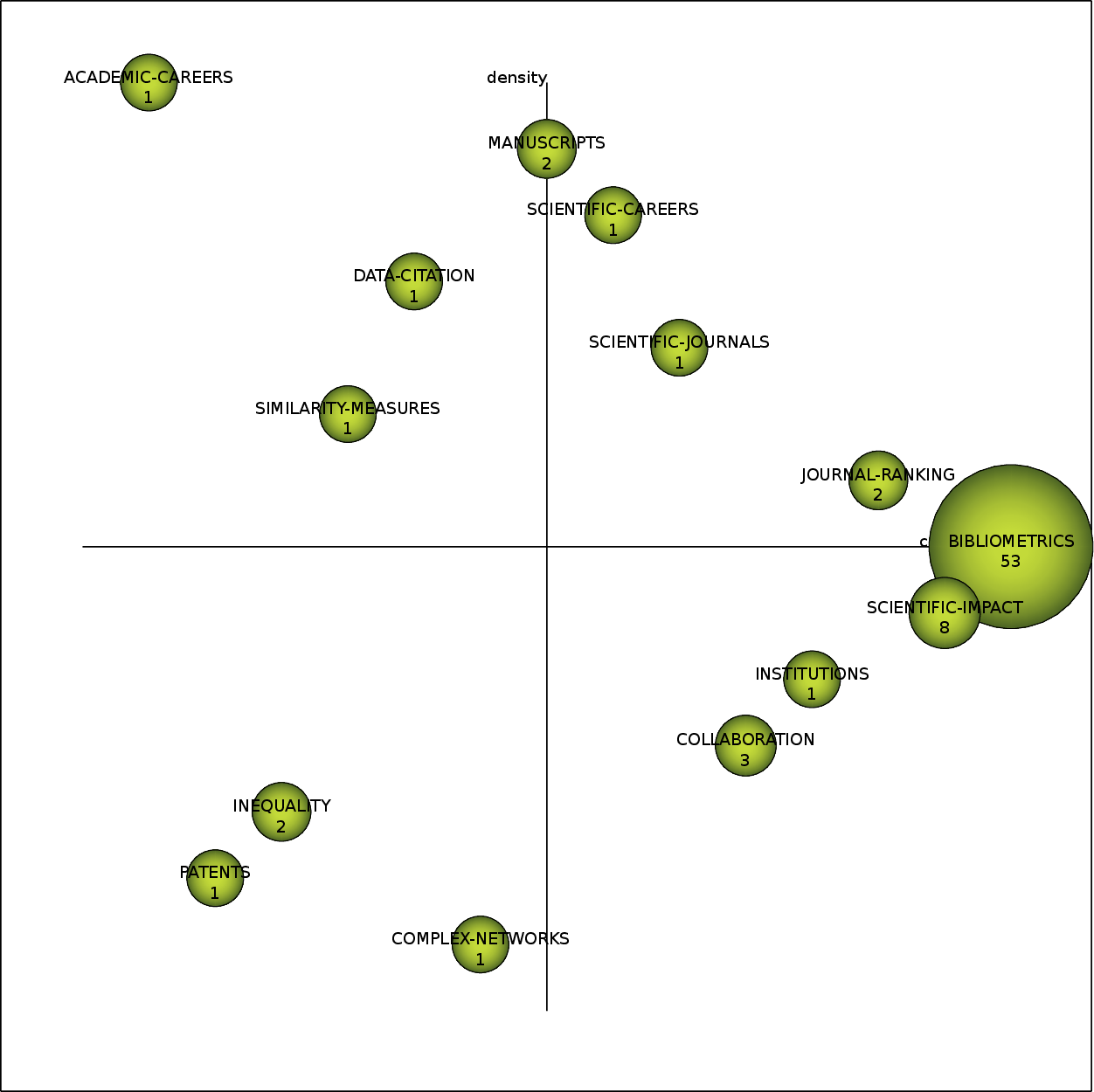}
  \caption{SciMAT strategic diagram: Period~3 (2019--2025).}
  \label{fig:scimat_sd_p2}
\end{figure}

The two approaches yield markedly different cluster counts across all periods. The proposed framework identifies 18, 12, and 9 clusters in Periods~1, 2, and~3, respectively, whereas SciMAT produces 7, 14, and 14 clusters over the same intervals (Table~\ref{tab:cluster_comparison}). This divergence reflects differences in community detection procedures -- the proposed framework relies on the Louvain algorithm applied to association-strength-normalised networks, while SciMAT implements single-pass centroid clustering -- as well as distinct thresholding strategies. Within the proposed framework, a minimum term frequency of five occurrences per period and an upper bound of 250 retained terms are imposed to limit sparsity while maintaining semantic coverage. The progressive reduction in cluster count is therefore consistent with increasing network cohesion and consolidation of thematic structure. In contrast, the expansion observed in SciMAT from 7 to 14 clusters appears closely associated with vocabulary growth across periods, leading to finer partitioning of the thematic space. The two approaches thus capture different aspects of structural evolution: consolidation under modularity-based community detection and fragmentation under centroid-based clustering. Figures~\ref{fig:scimat_sd_p0}--\ref{fig:scimat_sd_p2} present the three SciMAT strategic diagrams, as exported from the corresponding HTML reports. In each diagram, nodes are positioned according to their centrality (horizontal axis) and density (vertical axis), computed relative to period-specific means that define the quadrant boundaries. Node size is proportional to the number of core documents associated with each theme.

\paragraph{Period~1 (2007--2012)} SciMAT identifies a single dominant \textit{citation} cluster (centrality~=~7.47, density~=~4.15; 25~core documents) positioned in the basic-themes quadrant, alongside six smaller and comparatively isolated clusters: \textit{patents}, \textit{citation-network}, \textit{informetrics}, \textit{power-law-models}, \textit{stochastic-model}, and
\textit{universities} (Figure~\ref{fig:scimat_sd_p0}). Three of these (\textit{informetrics}, \textit{power-law-models}, \textit{stochastic-model}) display centrality values approaching zero and occupy peripheral or niche positions. In the same period, the proposed framework identifies 18~clusters,
differentiating among \textit{h-index}, \textit{citation analysis}, \textit{citation}, \textit{bibliometrics}, and \textit{scientific collaboration}
as distinct yet structurally interconnected thematic nodes. The allocation of \textit{h-index} to the motor-themes quadrant and \textit{citation} to the basic-themes quadrant introduces a clearer articulation between internally cohesive indicator research and transversal citation-based constructs. By comparison, SciMAT's single \textit{citation} cluster aggregates several of these components within a unified theme, resulting in a more compact but less differentiated
representation of the period's evaluative core.

\paragraph{Period~2 (2013--2018)} SciMAT produces 14~clusters, with \textit{bibliometrics} emerging as the dominant node (centrality~=~19.66,
density~=~4.45; 52~core documents). One structural feature of the partition concerns the placement of \textit{impact} (centrality~=~5.37, density~=~25.0) in the motor-themes quadrant despite containing only 2~core documents (Figure~\ref{fig:scimat_sd_p1}). This configuration is consistent with the well-known sensitivity of density measures to very small clusters, where limited internal links can produce disproportionately high density values. Several additional clusters -- \textit{database-errors}, \textit{patent-citations}, \textit{scientists}, \textit{citation-contexts}, \textit{collaboration-networks}, and \textit{citation-practices} -- exhibit centrality values below~1.5 and are
based on a single core document, suggesting highly localised thematic
segments with limited integrative capacity. Notably, SciMAT does not extract \textit{altmetrics} as a distinct named cluster in this period. In contrast, the proposed framework identifies \textit{altmetrics} in the lower-left quadrant (61 occurrences), capturing its emergence as a recognisable thematic trajectory prior to full structural consolidation.

\paragraph{Period~3 (2019--2025)} SciMAT again generates 14~clusters, dominated
by \textit{bibliometrics} (centrality~=~27.66; 53~core documents) and \textit{scientific-impact} (centrality~=~11.66; 8~core documents)(Figure~\ref{fig:scimat_sd_p2}). The remaining clusters are primarily peripheral, with 10 of the 12 additional themes exhibiting centrality values below 4 and single-document cores, indicating limited structural integration within the network. Two configurations identified by the proposed framework in this period -- \textit{machine learning and artificial intelligence} and \textit{science of science} -- do not emerge as consolidated structures in the SciMAT partition. Their absence as distinct clusters is plausibly related to the dispersion of their constituent vocabulary across multiple SciMAT themes, without sufficient centroid proximity to produce a unified cluster under single-pass centroid clustering.

\begin{figure}[ht]
  \centering
  \includegraphics[width=0.55\textwidth]{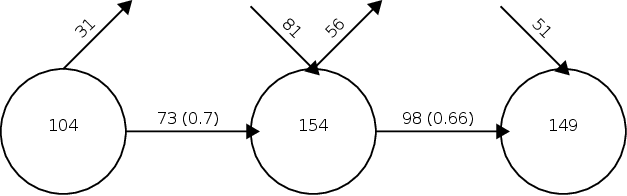}
  \caption{SciMAT overlapping map. Circles represent the three
    sub-periods (2007--2012: 104~terms; 2013--2018: 154~terms;
    2019--2025: 149~terms). Numbers on arrows denote shared terms
    between adjacent periods; values in parentheses indicate the
    corresponding normalised overlap index.}
  \label{fig:scimat_overlap}
\end{figure}

\begin{figure}[!ht]
  \centering
  \includegraphics[width=0.6\textwidth]{%
    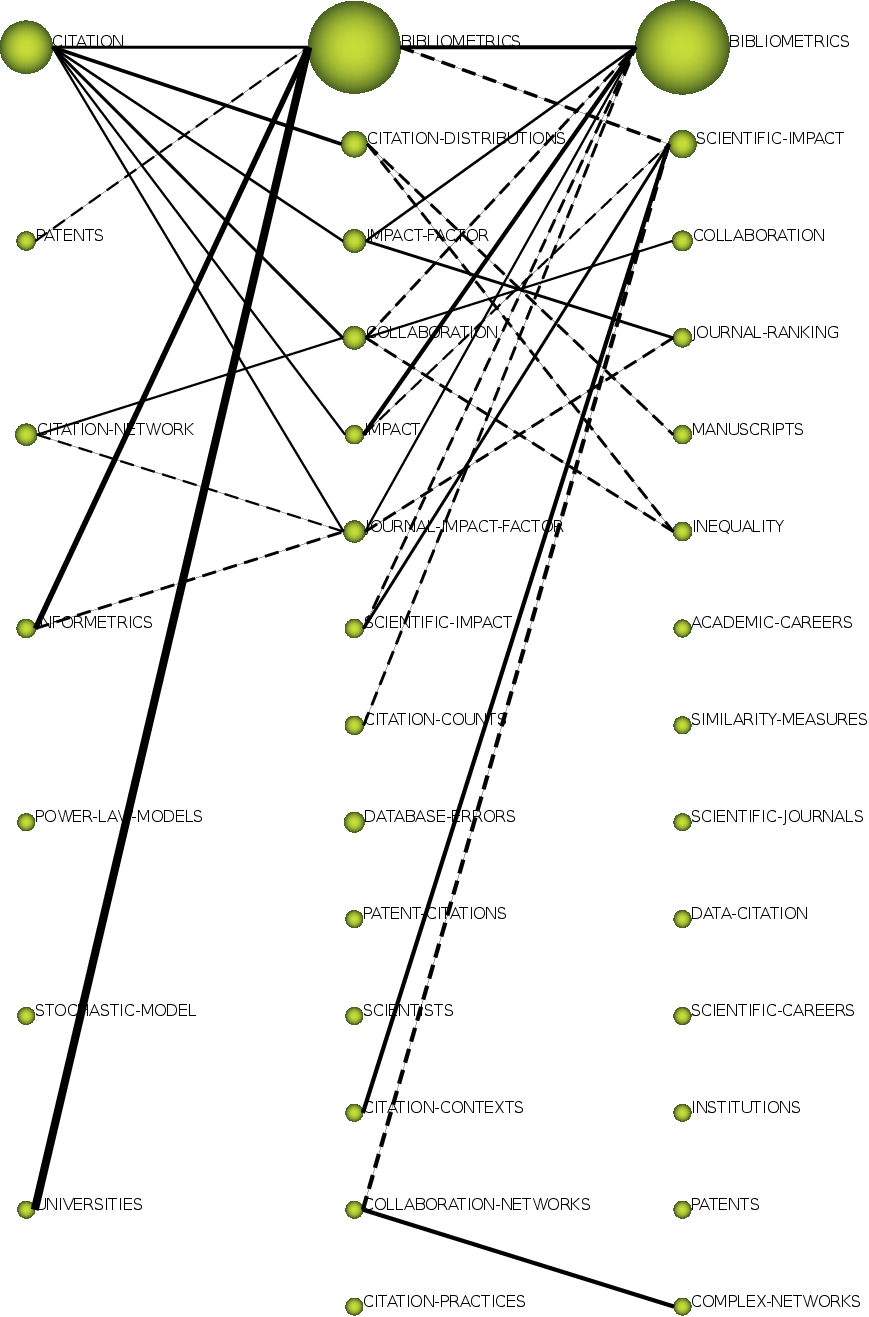}
  \caption{SciMAT evolution map based on the inclusion index
    (core-documents count). Node area is proportional to the number of
    core documents; edge thickness reflects inclusion weight. Strong
    links (solid) correspond to above-threshold inclusion; weak links
    (dashed) are below-threshold connections retained by the algorithm.}
  \label{fig:scimat_evol}
\end{figure}

\paragraph{Inter-temporal lineage} The overlapping map (Figure~\ref{fig:scimat_overlap}) and the evolution map (Figure~\ref{fig:scimat_evol}) summarise SciMAT's reconstruction of longitudinal dynamics. The overlapping map reports the number of shared terms between adjacent periods: 73 terms overlap between Periods 1 and 2 (Jaccard-type index = 0.70), and 98 terms between Periods 2 and 3 (index = 0.66), suggesting substantial vocabulary continuity over time. The evolution map, based on core documents' inclusion indices, displays a pronounced hub-and-spoke configuration centred on
\textit{bibliometrics}.

From Period~1 to 2, the dominant \textit{citation} cluster distributes lineage across \textit{bibliometrics}~(3.57), \textit{citation-distributions}~(4.29), \textit{impact-factor}~(2.86), \textit{collaboration}~(3.33), \textit{impact}~(2.50), and \textit{journal-impact-factor}~(2.50). Together,
\textit{informetrics} and \textit{universities} connect to \textit{bibliometrics} with inclusion weights of 6.67 and 10.00, respectively. These values primarily reflect lexical inclusion within core-document sets, as \textit{universities} contributes exclusively to \textit{bibliometrics} despite constituting a peripheral, single-document cluster. From Period~2 to 3, \textit{bibliometrics} absorbs most of the lineage mass: \textit{impact}~$\to$~\textit{bibliometrics}~(5.00),
\textit{impact-factor}~$\to$~\textit{bibliometrics}~(2.86), \textit{collaboration}~$\to$~\textit{bibliometrics}~(3.33), and \textit{scientific-impact}~$\to$~\textit{bibliometrics}~(3.33). Only three connections extend to Period~2 clusters beyond \textit{bibliometrics}: \textit{impact-factor}~$\to$~\textit{journal-ranking}~(3.33), \textit{collaboration}~$\to$~\textit{collaboration}~(2.50), and \textit{scientific-impact}~$\to$~\textit{scientific-impact}~(3.33). The resulting structure is strongly centralised around a single dominant hub.

By comparison, the proposed framework reconstructs a more articulated evolutionary configuration. The lineage strength measure, integrating PageRank-weighted term inclusion and mutual structural relevance, captures both strong continuations and selective transmissions across periods. The differentiation of the
\textit{citation} cluster from Period~1 into \textit{h-index}, \textit{citation analysis}, and \textit{altmetrics} in Period~2 emerges through distinct lineage strengths reflecting differential coverage and structural salience. Similarly, the convergence of \textit{collaboration}, \textit{citation network}, and \textit{h-index} in Period~2 towards the \textit{science of science} cluster in Period~3 is reconstructed as the merger of structurally related streams. This pattern does not appear in the SciMAT evolution map, which is driven by inclusion relations over core-document sets and therefore tends to privilege dominant lexical intersections. The gradual contraction of \textit{h-index} across periods, observable through declining fuzzy cardinality in the proposed framework, is likewise not separately visible in SciMAT's output, where it is absorbed within the
\textit{bibliometrics} trajectory at successive transitions.

The results shown here indicate that the differences between the two strategies are not merely quantitative variations in cluster counts or lineage weights, but stem from distinct modelling assumptions about what constitutes thematic continuity. Whereas inclusion-based methods reconstruct evolution through lexical intersection among core-document sets, the proposed framework embeds lineage within the weighted relational structure of clusters and incorporates graded document affiliation into the linkage mechanism. The consequent divergence in evolutionary topology -- centralised hub formation vs articulated split-and-merge patterns -- reveals how alternative operationalisations of continuity yield substantively different representations of knowledge dynamics. These findings encourage broader reflection on the conceptual foundations of longitudinal science mapping and the implications of structurally integrated modelling for the interpretation of thematic change.

\section{Discussion}\label{sec:5}
The present study advances a structural reformulation of thematic evolution within science mapping by resolving a methodological inconsistency embedded in classical longitudinal implementations. In established approaches, themes are detected through relational clustering within co-occurrence networks, yet their evolution is reconstructed through set-theoretic overlap among core documents or keyword lists. This dual logic implicitly shifts the ontological status of themes from network-embedded relational configurations to bounded lexical sets when moving from cross-sectional to longitudinal analysis. The framework proposed here restores coherence by embedding lineage reconstruction within the same weighted relational architecture that underpins cross-sectional detection.

Conceptually, this integration reframes thematic continuity as a structurally mediated process. Continuity is no longer inferred solely from shared vocabulary, but from the transformation of weighted configurations in which term salience and document affiliation are jointly considered. By incorporating PageRank-weighted term importance and fuzzy document membership, the model captures graded and directional continuity, distinguishing between broad lexical coverage and mutual structural relevance. This distinction enables the analysis of thematic change by examining how intellectual structures are differentially retained, reweighted, or reconfigured across periods. The resulting representation treats thematic change as a dynamic reconfiguration of relational structures rather than as persistence of surface lexical overlap. Methodologically, three elements constitute the framework's core innovation. First, fuzzy thematic affiliation replaces exclusive publication assignment with graded membership. Scientific contributions often span multiple thematic strands, and binary partitioning may obscure this multiplicity. The adoption of fuzzy membership provides a more faithful representation of thematic embeddedness and permits the estimation of cluster size via fuzzy cardinality, capturing both the intensity and the dispersion of participation. Second, lineage strength is decomposed into complementary components -- weighted inclusion and structural importance -- thereby separating directional coverage from reciprocal salience. This decomposition enhances interpretability and renders explicit the modelling assumptions underlying continuity measurement. Third, the parameterisation of lineage weighting through $\alpha$ introduces analytical transparency. Rather than embedding weighting choices implicitly within an index, the framework exposes them as tunable components, facilitating methodological scrutiny and replication.

The comparative assessment illustrates that alternative operationalisations of continuity yield substantively different evolutionary topologies. Inclusion-based methods privilege dominant lexical intersections and tend to generate centralised hub configurations, in which major themes absorb lineage mass across transitions. By contrast, a structurally integrated specification reveals articulated split-and-merge patterns, as well as gradual differentiation or attenuation of thematic strands. These differences are not merely quantitative, reflecting distinct epistemic commitments about what constitutes thematic persistence. Beyond its technical modifications, the proposed framework offers a more explicit articulation of the conceptual premises that inform longitudinal science mapping. At the same time, the scope of the contribution is shaped by several methodological contingencies. The resulting configurations remain dependent on clustering decisions, as partition boundaries are inherently resolution-sensitive and alternative community detection algorithms may yield different yet equally defensible structures. The importance index, in turn, relies on PageRank as a proxy for semantic salience within clusters; although this measure captures recursive centrality and attenuates degree bias, other centrality specifications could foreground different structural properties. Network topology is also influenced by preprocessing choices, such as frequency thresholds and lexical harmonisation, which, despite being explicitly documented, introduce degrees of analytical discretion that warrant transparent reporting and, where possible, sensitivity assessment. Finally, the reliance on discrete temporal segmentation -- while consistent with established strategic mapping practice -- necessarily imposes boundaries on processes of thematic change that unfold more continuously in practice. These considerations do not undermine the framework's contribution but clarify its domain of validity. The proposed integration enhances structural coherence between cross-sectional and longitudinal modelling, increases interpretive transparency, and reduces artefacts arising from purely lexical lineage reconstruction. At the same time, it remains embedded within the broader assumptions of network-based science mapping and shares their interpretative and methodological constraints, intrinsic to clustering and keyword-based analysis.

\section{Conclusions and Final Remarks}\label{sec:6}
This study advances a structurally integrated framework for modelling thematic evolution in science mapping. The point of departure is a methodological asymmetry in classical longitudinal approaches, which combine relational clustering for cross-sectional detection with set-theoretic criteria for inter-temporal linkage. By situating both detection and lineage reconstruction within a unified weighted network architecture and by introducing graded document affiliation, the framework restores coherence between theme identification and the modelling of temporal transformation. The central contribution is to redefine thematic continuity as a structurally grounded process: evolution is reconstructed through weighted relational configurations that jointly incorporate term salience and document-level participation. In this way, the framework preserves the interpretability of strategic diagrams -- valued for their capacity to position themes along centrality and density dimensions -- while extending their analytical reach to longitudinal dynamics. Rather than tracking vocabulary alone, the model captures how intellectual structures are progressively reconfigured, consolidated, or redistributed across periods.

Beyond the empirical application presented here, the framework reinforces the methodological foundations of longitudinal science mapping by making explicit the assumptions underlying lineage construction and by formalising the balance between directional coverage and structural relevance. Such explicit parameterisation enhances analytical transparency, facilitates replication, and enables systematic comparison across studies without sacrificing interpretability. In doing so, the study situates science mapping within a broader network-analytic perspective, where consistency between structural detection and dynamic modelling is regarded as essential for coherent temporal analysis.

More generally, the framework prompts reconsideration of the epistemic status of themes in bibliometric research. When themes are understood as evolving relational configurations rather than as bounded lexical aggregates, longitudinal analysis shifts from tracking stable labels to modelling structural transformation. This reorientation contributes to the conceptual consolidation of thematic analysis and positions science mapping as a dynamically consistent approach within the broader field of knowledge evolution studies. Further progress in modelling thematic change is likely to depend less on incremental refinements of similarity indices and more on the development of structurally coherent dynamic models capable of accommodating the complexity of intellectual transformation. In this perspective, the present framework offers a principled basis for subsequent theoretical refinement and empirical deployment. Promising directions include the integration of thematic networks with additional relational layers -- such as citation, authorship, or institutional ties -- so as to capture multidimensional evolutionary processes within multiplex structures; the development of adaptive weighting strategies for lineage strength, in which the balance between inclusion and structural relevance is empirically calibrated; and the exploration of overlapping or rolling temporal windows to approximate continuous thematic drift while preserving the interpretability of strategic mapping.

\bibliographystyle{abbrvnat}
\bibliography{biblio}

\end{document}